# Computational Search for Magnetic and Non-magnetic 2D Topological Materials using Unified Spin-orbit Spillage Screening

Kamal Choudhary[1], Kevin F. Garrity[1], Jie Jiang[2], Ruth Pachter[2], Francesca Tavazza[1]

1 Materials Science and Engineering Division, National Institute of Standards and Technology, Gaithersburg, Maryland 20899, USA.

2 Materials Directorate, Air Force Research Laboratory, Wright–Patterson Air Force Base, Ohio 45433, USA.

**Abstract**

Two-dimensional topological materials (TMs) have a variety of properties that make them attractive for applications including spintronics and quantum computation. However, there are only a few such experimentally known materials. To help discover new 2D TMs, we develop a unified and computationally inexpensive approach to identify magnetic and non-magnetic 2D TMs, including gapped and semi-metallic topological classifications, in a high-throughput way using density functional theory-based spin-orbit spillage, Wannier-interpolation, and related techniques. We first compute the spin-orbit spillage for the ~1000 2D materials in the JARVIS-DFT dataset (https://www.ctcms.nist.gov/~knc6/JVASP.html ), resulting in 122 materials with high-spillage values. Then, we use Wannier-interpolation to carry-out $Z_2$, Chern-number, anomalous Hall conductivity, Curie temperature, and edge state calculations to further support the predictions. We identify various topologically non-trivial classes such as quantum spin-hall insulators (QSHI), quantum anomalous-hall insulators (QAHI), and semimetals. For a few predicted materials, we run $G_0W_0$+SOC and DFT+U calculations. We find that as we introduce many-body effects, only a few materials retain non-trivial band-topology, suggesting the importance of high-level DFT methods in predicting 2D topological materials. However, as an initial step, the automated spillage screening and Wannier-approach provide useful predictions for finding new topological materials and to narrow down candidates for experimental synthesis and characterization.

**Corresponding author:** Kamal Choudhary (E-mail: kamal.choudhary@nist.gov)

# 1 Introduction

In recent years, there has been a huge upsurge in topological materials research, following the predictions and observations of Dirac, Weyl, and Majorana fermions in condensed matter systems[1,2]. Several classes of topological materials have been proposed for applications in error-reduced quantum computing[3-6], or as high mobility and dissipationless conductors. While there have been several recent detailed screening efforts for 3D non-magnetic topological materials[7-17], such systematic searches for 2D materials are still developing[18-20], especially for magnetic systems. Nevertheless, 2D materials could be more important than 3D ones because of their unique potential as miniaturized devices and their tunability via layering or functionalization[21].

2D topological insulating phases can be classified into two primary types: quantum spin Hall insulators (QSHI[22]), which have time-reversal symmetry (TRS), and quantum anomalous Hall insulators (QAHI[23,24]), which lack TRS. QSHIs, characterized by a $Z_2$ invariant, have an insulating bulk and Dirac cone edge features. Examples include graphene, silicene[25], germanene[26], stanine[27], and 1T' metal dichalcogenides[28,29]. Quantum anomalous Hall insulators (QAHI), characterized by a Z invariant known as the Chern number, are magnetic materials with a bulk gap and quantized conducting edge channels, even in the absence of an external magnetic field. Thus far, QAHI-like behavior has been observed experimentally in very few systems: Cr,V doped $(Bi,Sb)_2Te_3$[23] and $MnBi_2Te_4$[30-33], under highly controlled conditions. While there have been many theoretical works predicting QAHIs in 2D materials, surfaces, or interfaces, a systematic investigation of monolayer 2D materials is lacking. Some previously explored examples include transition metal halides such as $CoBr_2$, $FeCl_3$, $NiRuCl_6$, $V_2O_3$, $FeCl_3$, $RuCl_3$[34-39].

Compared to gapped topological systems, semimetals in two-dimensions with spin-orbit coupling are relatively under-explored. For example, graphene is a well-known example of a two-dimensional semimetal; however, the Dirac point in graphene is not robust to the addition of a finite amount of spin-orbit coupling[22]. This splitting of symmetry-protected band crossings by spin-orbit coupling occurs for most crossings protected by two-dimensional point group symmetries. Also, unlike in three dimensions, where Weyl points can occur at generic points in the Brillion zone, Weyl points will not occur generically in two dimensions without the tuning of some external parameter[40]. However, it is possible for non-symmorphic symmetries to protect Dirac or Weyl points in two-dimensions[41], and we find several examples of this in our work. While 2D insulting topological materials are relatively easy to classify using topological indices such as $Z_2$ and the Chern number, finding topological semimetals is non-trivial. Hence, having a universal strategy for screening both topological insulators and semimetals is highly desirable.

In our previous work[8], we discovered several classes of non-magnetic topological 3D materials using spin-orbit spillage[42] to perform the initial screening step. The spillage technique is based on comparing density functional theory (DFT) wave-functions obtained with- and without spin-orbit coupling. Materials with high spillage value (discussed later) are considered non-trivial. In this paper, we extend this approach to screen the JARVIS-DFT 2D database to search for topological insulators and semimetals, with and without TRS. The JARVIS-DFT database is a part of Materials Genome Initiative (MGI) at National Institute of Standards and Technology (NIST) and contains about 40000 3D and 1000 2D materials with their DFT-computed structural, energetics[43], elastic[44], optoelectronic[45], thermoelectric[46], piezoelectric, dielectric, infrared[47], solar-efficiency[48,49], and topological[8] properties. Most of the 2D crystal structures we consider are derived from structures in the experimental Inorganic Crystal Structure Database (ICSD) database[22], implying that most of them should be experimentally synthesizable. In this work, we screen materials from the JARVIS-DFT 2D database, searching for low bandgap materials with high-spillage[8] values, identifying candidate 2D TMs. Then, for insulating compounds, we systematically carry out $Z_2$ calculations for non-magnetic materials and Chern number calculations for magnetic materials. For metals, we search for band crossings near the Fermi level. We also predict surface (2D) and edge (1D) band-structures, Curie temperatures, and anomalous Hall conductivity. For a subset of predicted materials, we run $G_0W_0$+SOC and DFT+U calculations. We find that as we introduce improved treatments of many-body effects, only a few materials retain a non-trivial band-topology, suggesting the importance of higher-level electronic structure methods in predicting 2D topological materials. However, as an initial step, the automated spillage screening and Wannier analysis provide useful predictions of potential topological materials, narrowing down candidate materials for experimental synthesis and characterization.

## 2 Results and discussion

As mentioned above, the spin-orbit spillage criterion is applicable for both magnetic/non-magnetic and metallic/insulating classes of materials. A flow-chart describing our computational search using the spillage as well as the traditional Wannier-based method is shown in Fig. 1. Starting from 963 2D materials in the JARVIS-DFT database, we first screen for materials with OptB88vdW bandgaps < 1.5 eV, because SOC-induced band-inversion is limited to the magnitude of SOC. This leads to 506 materials. Then we carry out spin-polarized calculations with and without SOC, and compute spillage using Eq.1. Setting a threshold of 0.5, we find 122 materials with high-spillage. Note that for magnetic materials, we also screen in terms of the magnetic moment, selecting only cases with magnetic moment > 0.5 $\mu_B$. As a computational note, for the spin-polarized calculations, we set the initial magnetic moment to a high initial value of 6 $\mu_B$ per atom, to search for high spin configurations. After this initial screening, we find 19 magnetic-insulating, 40 magnetic-metallic, 10 non-magnetic insulating, and 53 non-magnetic metallic materials. In Table. 1, we present selected examples of each class of topological materials that we consider. The full list is given in the supplementary information. Clearly, metallic topological candidates outnumber insulators. Note that previous 2D topological material searches were mostly limited to insulators, but our approach can be extended to semimetals as well. Some of the predicted materials have already been experimentally synthesized, including $Bi_2TeI$[50], $RuCl_3$[51], $FeTe$[52] etc., but the experimental confirmation of their topological properties is still ongoing.

In the remaining part of the discussion, we analyze the overall distribution of topological materials, and then we focus on individual topological classes, exploring a few examples. Information for other similar materials is provided in the JAVRIS-DFT database. Finally, we discuss a few cases of DFT+U and $G_0W_0$+SOC calculations as a way to understand the limits of semi-local DFT calculations as screening tools for topological properties.

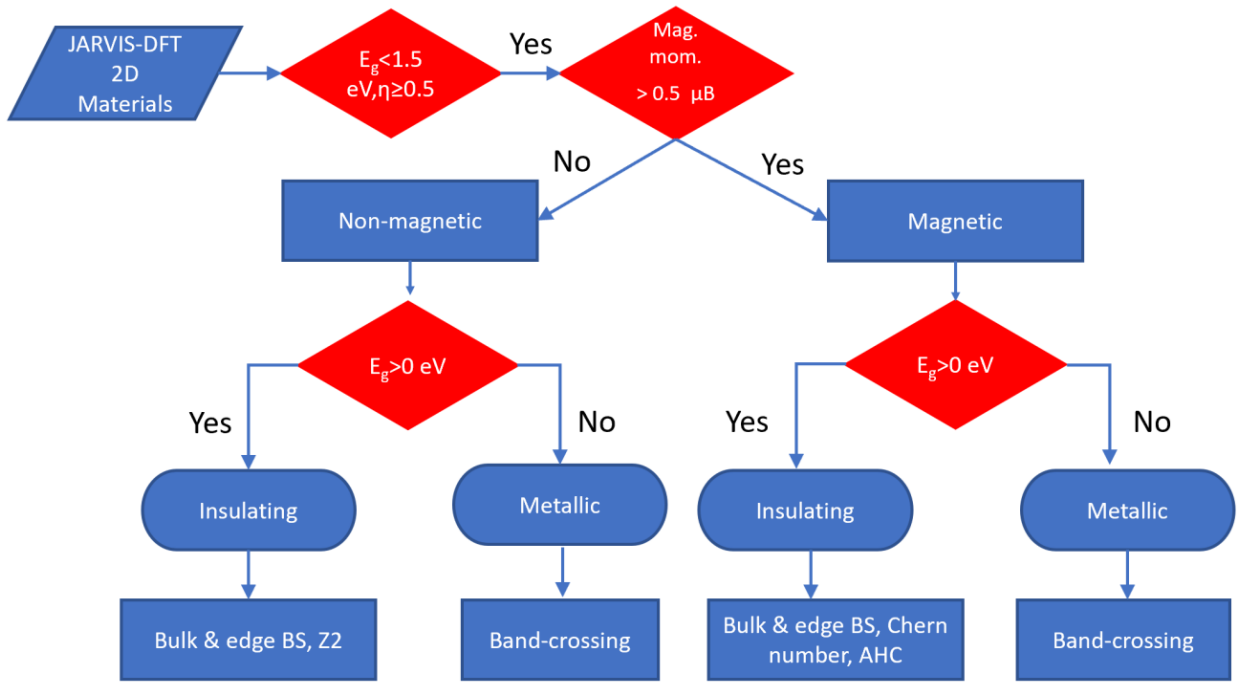

*Fig. 1 Flow-chart showing the screening and analysis methodology.*

## 2.1 Spin-orbit spillage analysis

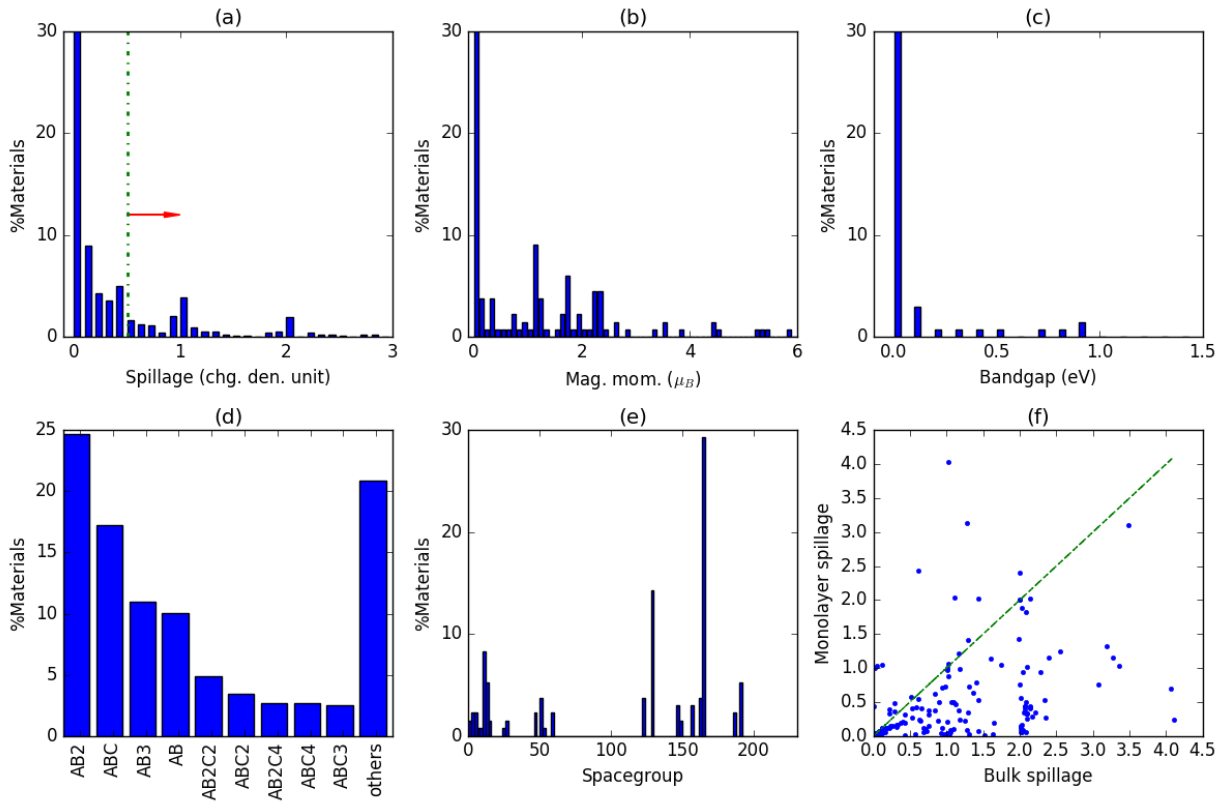

*Fig. 2 Spillage assisted screening of 2D topological materials in JARVIS-DFT database. a) spillage distribution of materials. Materials with spillage ≥ 0.5 are likely to be topological, b) magnetic moment distribution for high-spillage materials, c) bandgap-distribution, d) chemical prototype distribution. Zero bandgap materials should be topological-semimetals., e) space-group distribution for high spillage materials, f) spillage of bulk (3D) vs monolayers (2D) for non-magnetic systems.*

In Figure 2 we show the spillage-based distribution for two-dimensional materials. We find that 122 materials have high spillage (with a threshold 0.5). The spillage criterion does, therefore, eliminate many materials, as shown in Fig. 2a. The 122 selected materials include both magnetic and non-magnetic materials, as well as both metals and insulators, and contains examples with a variety of chemical prototypes and crystal structures. The magnetic moments of topological materials range from 0 to 6 µB, as shown in Fig. 2b. The most common topological 2D material types are $AB_2$, ABC, $AB_3$, and AB. Note that in experimental synthesis it might be easier to synthesize simple chemical prototypes such as AB or 1:1 structures. Hence, having a variety of chemical prototypes provides a vast amount of opportunity for synthesis. We find that most of the high-spillage materials belong to highly symmetric space groups, as shown in Fig. 2e. This is consistent with our previous 3D topological materials search. A comparison of 3D and 2D spillage in 2f indicates that the 2D-monolayer spillage values are generally lower than their 3D-bulk counterparts, but the trend is rather weak. This may be explained in part by quantum confinement in 2D-monolayers tending to increase band gaps.

*Table 1. Examples of various classes of 2D topological materials using spillage ($\eta$) and Wannier calculations. JID represents the JARVIS-ID, Spg. the spacegroup, $E_g$ the bandgap (eV), $E_f$ the exfoliation energy (meV/atom), $\eta$ the spillage, TopoClass the topological class of the material, $T_c$ the Curie temperature. Here, QSHI, QAHI, NM-Semi, M-Semi represents quantum spin Hall insulators, quantum anomalous Hall insulators, non-magnetic semimetals and magnetic semimetals.*

| Formula | JID | Spg. | $E_g$ (eV) | $\eta$ | $E_f$ (meV/atom) | TopoClass | $T_c$ (K) |
|---|---|---|---|---|---|---|---|
| **$Bi_2TeI$** | 6901 | $P\bar{3}m1$ | 0.053 | 3.1 | 43.9 | QSHI | - |
| **BiI** | 6955 | C2/m | 0.145 | 3.1 | 77.7 | QSHI | - |
| **$HfTe_5$** | 19987 | Pmmn | 0.074 | 0.61 | 88.3 | QSHI | - |
| **$TaIrTe_2$** | 6238 | $P2_1/m$ | 0.034 | 1.02 | 100.4 | QSHI | - |
| **$ZrFeCl_6$** | 13600 | P312 | 0.011 | 1.01 | 72.0 | QAHI | 5.3 |
| **$CoBr_2$** | 6034 | $P\bar{3}m1$ | 0.019 | 1.00 | 75.5 | QAHI | 17.7 |
| **$VAg(PSe_3)_2$** | 60525 | C2 | 0.018 | 0.50 | 74.7 | QAHI | - |
| **Ti$_2$Te$_2$P** | 27864 | $P\bar{3}m1$ | 0.0 | 1.06 | 58.9 | NM-Semi | - |
| **ZrTiSe$_4$** | 27780 | P2/m | 0.0 | 2.00 | 96.1 | NM-Semi | - |
| **MoS$_2$** | 730 | $R\bar{3}m$ | 0.0 | 0.98 | 89.4 | NM-Semi | - |
| **AuTe$_2$** | 27775 | $P\bar{3}m1$ | 0.0 | 0.50 | 317.6 | NM-Semi | - |
| **MnSe** | 14431 | P4/nmm | 0.0 | 0.82 | 81.8 | M-Semi | -- |
| **FePSe$_3$** | 27940 | $P\bar{3}m1$ | 0 | 1.07 | 74.9 | M-Semi | 148.3 |
| **FeTe** | 6667 | P4/nmm | 0.0 | 1.2 | 90.2 | M-Semi | 224.7 |
| **CoO$_2$** | 31379 | C2/m | 0.0 | 2.01 | - | M-Semi | - |
| **TiCl$_3$** | 13632 | $P\bar{3}m1$ | 0.0 | 0.62 | 72.4 | M-Semi | -- |
| **VBr$_2$O** | 6832 | Pmmm | 0.0 | 0.45 | 67.7 | M-Semi | - |
| **Co (OH)$_2$** | 28106 | C2/m | 0.0 | 2.1 | - | M-Semi | -- |

## 2.2 Individual case studies

### 2.2.1 Quantum Spin Hall Insulator (QSHI)

Some of the semiconducting non-magnetic materials with high-spillage that we identified are: BiI (JVASP-6955), TiI$_3$ (JVASP-6118), TaIrTe$_2$ (JVASP-6238), Bi$_2$TeI (JVASP-6901) and Bi (JVASP-20002). Materials such as 2D-Bi[53] are experimentally known as QSHI. The fact that we find them using the spillage criteria further supports the applicability of this method to 2D materials. In order to computationally confirm that they are QSHI, we calculate the Z$_2$ index for a few candidates (BiI, TiI$_3$, Bi, Bi$_2$TeI). We find these materials indeed have non-zero indices. We show examples of high spillage QSHI in Fig. 3. We observe high spillage peaks for all three materials (Fig. 3a,d,g), which occur near the Z and Gamma points. It is difficult to observe band-inversion in Fig. c because all the states near $E_f$ are derived from Bi orbitals, so the orbital-projected bandstructure doesn't show any obvious band-inversion. However, the spillage method suggests such inversion as in Fig. 3e. In all of these cases, we find that the bulk (Fig. 2b,e,h) is insulating while the edge states (Fig. 2c,f,i) are conducting, with Dirac dispersions at appropriate high symmetry points, as expected for QSHI.

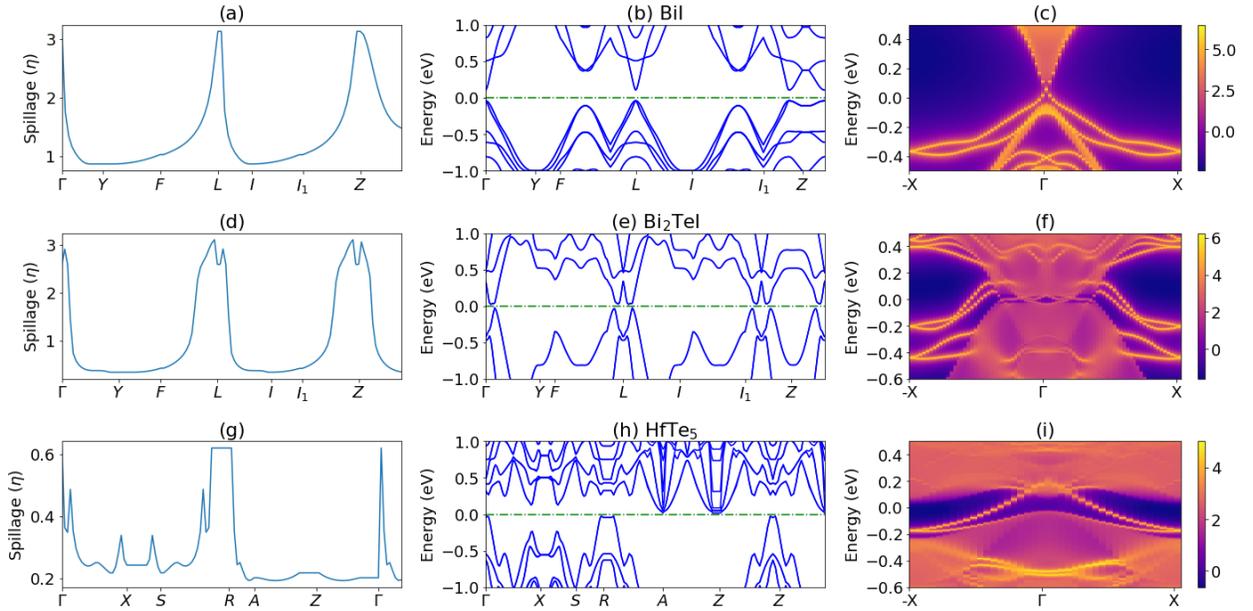

*Fig. 3 Examples of quantum spin Hall insulators (QSHI) with spillage (a,d,g), surface (b,e,h), and edge bandstructure (c,f,i) plots. (a,b,c) for BiI, (d,e,f) for $Bi_2TeI$, (g,h,i) for $HfTe_5$. In (c,f,i) the color scale represent the electron occupation. The Fermi energy is set to zero in all band structure plots.*

## 2.2.2 Quantum Anomalous Hall Insulator (QAHI)

Next, we investigate the candidates for QAHI materials. We have 19 such candidates after screening for spillage > 0.5, bandgap > 0.0, magnetic moment > 0.5. We carry out Chern number calculations, finding several QAHI materials. Note that these were obtained using DFT calculations only, i.e. without DFT+U or $G_0W_0$ corrections; such corrections will be discussed later. Some of the materials with non-zero Chern numbers are: $CoCl_2$ (JVASP-8915), $VAg(PSe_3)_2$ (JVASP-60525), $ZrFeCl_6$ (JVASP-13600). We show the surface and edge band-structures as well as anomalous Hall conductivity (AHC) in Fig 4 for $VAg(PSe_3)_2$ and $ZrFeCl_6$. They both have Chern number equal to 1. Correspondingly, there is a single edge channel that connects the valence band (VB) to the conduction band (CB) (Fig 4 c, d), which carries the quantized AHC (Fig 4 d, h). Like QSHI, QAHI materials have gapped surfaces (Fig 4 b, f) but conducting edges.

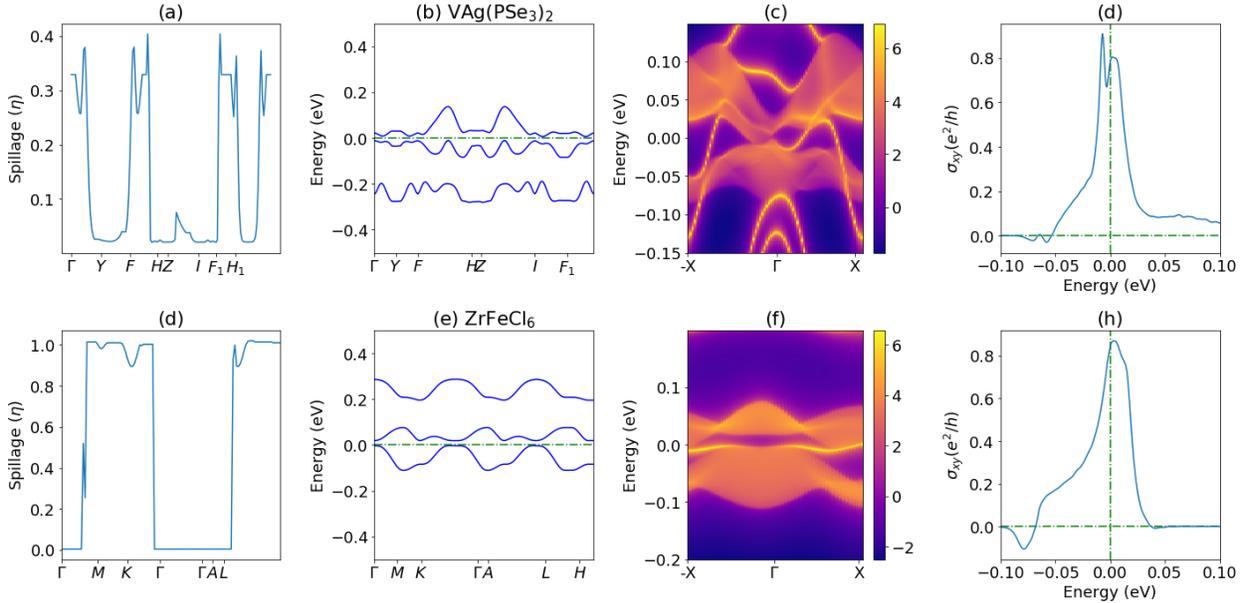

*Fig. 4 Examples of quantum anomalous Hall insulators (QAHI) with spillage (a,e), surface (b,f), and edge bandstructure (c,g), and anomalous Hall conductivity (AHC) (d,h) plots. (a,b,c,d) for $VAg(PSe_3)_2$, (e,f,g,h) for $ZrFeCl_6$. The Fermi energy is set to zero in all band structure plots.*

### 2.2.3 Non-magnetic semimetals

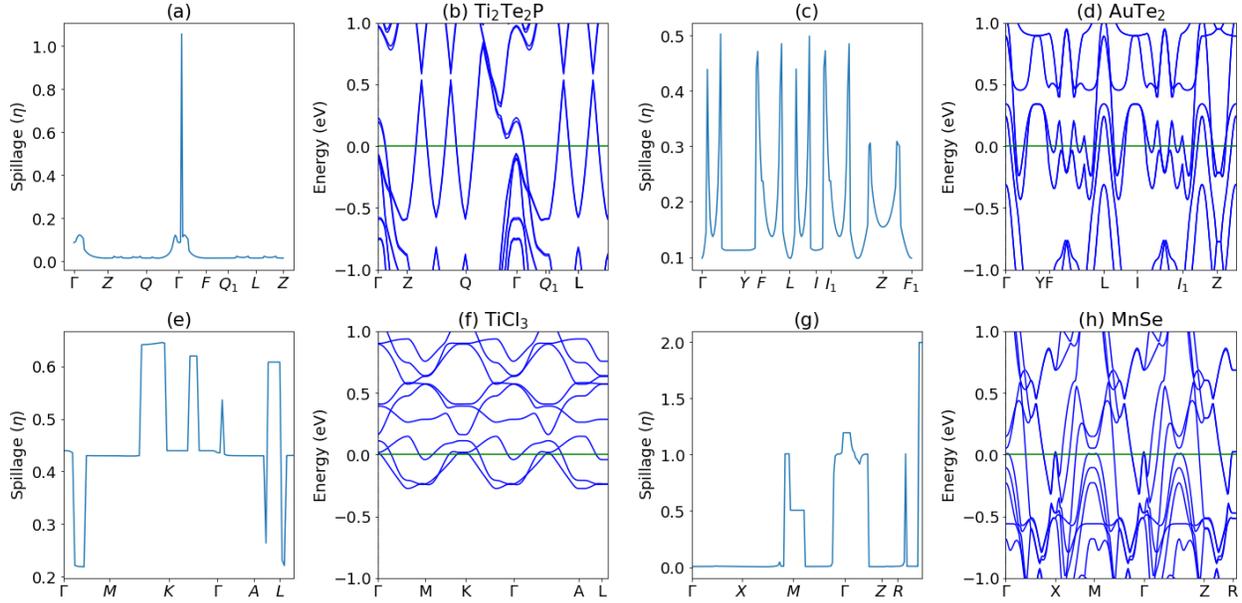

*Fig. 5 Examples of non-magnetic (a,b,c,d) and magnetic (e,f,g,h) topological semimetals with spillage (a,c,e,g) and surface bandstructure (b,d,f,g) plots. a-b) Ti$_2$TeP, c-d) AuTe$_2$ e-f) TiCl$_3$ g-h) MnSe*

Next, we identify the non-magnetic topological semimetals. This is more challenging than analyzing insulators because there are no specific $Z_2$/Chern-like indices for such materials. Rather we look for Weyl/Dirac nodes between the nominal valence and conduction bands. Examples of semimetals we found and confirmed with Wannier based methods are MoS$_2$ (JVASP-730), AuTe$_2$ (JVASP-27775), Ti$_2$Te$_2$P (JVASP-27864) and ZrTiSe$_4$ (JVASP-27780). The complete list of found topological semimetals is given in the supplementary information section, together with their spillage values. As is evident from Fig. 5 a and c, the spillage plots have spiky peaks that represent band-inversion. For Ti$_2$Te$_2$P the band-crossing is slightly below the Fermi level while for ZrTiSe$_4$ it is very close to the Fermi level. It is often possible to shift the Fermi-level of 2D materials via electrostatic or chemical doping, so band-crossings slightly away from Fermi-level can still have observable effects.

### 2.2.4 Magnetic semimetals

Finally, we look at ferromagnetic phases with band crossings, which are all Weyl crossings. We find band crossings near the Fermi level in MnSe (JVASP-14431), FeTe (JVASP-6667), $CoO_2$ (JVASP-31379), $TaFeTe_3$ (JVASP-60603), $TiCl_3$ (JVASP-13632), $Co(HO)_2$ (JVASP-28106) all of which have non-symmorphic symmetries that can protect band crossings[41]. Also, we find crossings in $VBr_2O$ (JVASP-6832), which has a double point group (*mmm*) that allows two-dimensional irreducible representations. In Fig 5e, we discuss $TiCl_3$ (JVASP-13632), a representative magnetic semimetal. The Weyl point occurs just above the Fermi level along the high symmetry line between Γ and M, as shown in Fig. 5f. Unfortunately, the valence band dips below the Fermi level at K, adding extra bands to the Fermi level. Similarly, in the case of MnSe the crossings are above the Fermi-level as shown in Fig. 5g and h. Note that the bandgaps and magnetic moments of the above systems are heavily dependent on the calculation method. This limitation of the methodology is discussed later.

### 2.3 Magnetic ordering

In addition to the electronic structure, we report the magnetic ordering of the structures and Curie temperature obtained using the method described in Ref.[54] model. Many of the examples we consider have either lower energy anti-ferromagnetic (AFM) phases or are ferromagnetic (FM) but with spins in-plane, but for the remaining materials, which we predict to be FM at finite temperature, we present their Curie temperatures and Chern numbers in Table 1. We find that $T_c$ of FeTe is unusually high because of the high *J* parameter resulting from the AFM configuration.

The FePS$_3$ and FeTe have $T_c$ higher than CrI$_3$, which suggests they can be used for above liquid nitrogen temperature $T_c$ topological applications. On the other hand, the $T_c$ of ZrFeCl$_6$ is very low, which is consistent with the large separation between Fe atoms in that crystal structure. As a next step, it would be interesting to study the layer dependence of $T_c$, which is beyond the scope of present work.

## 2.4 Limitations of GGA-based Calculations

Next, we investigate the reliability of our calculations for a few systems using GGA+U[55,56] and G$_0$W$_0$+SOC[57-59] methods. We note that fully *ab initio* beyond-DFT methods like G$_0$W$_0$ are very time-consuming, and hence unfeasible for a high-throughput search, while GGA+U is computationally fast, but has an adjustable parameter that limits predictive power. In Fig. 6a we show the GGA+U dependence of bandgaps and Chern numbers of ZrFeCl$_6$. We find that as we increase the U parameter, ZrFeCl$_6$ has a non-zero Chern number only for U < 0.2. This indicates that Chern behavior can be quite sensitive to U. Similar behavior was also observed for FeX$_3$ (X=Cl,Br,I) by Li[35], where the critical value of U varies from 0.43 to 0.80 eV. In Fig. 6b we show how G$_0$W$_0$+SOC can change the band gaps for VAgP(Se$_3$)$_2$, an example QAHI. In this case, G$_0$W$_0$+SOC increases the bandgap (0.018 eV vs 0.103 eV), but the band-shapes remain very similar, and the topology is non-trivial even in the G$_0$W$_0$+SOC calculation. However, for the other cases, we found that the band gap increase was substantial (Table. 2), and caused the bands to become un-inverted, resulting in trivial materials. We note that G$_0$W$_0$+SOC can also predict incorrect band-structures, especially for correlated transition metal compounds[60]. Here we are using only single-shot GW (G$_0$W$_0$), and more accurate results could be obtained by using fully self-consistent (sc)-GW[63], which is not carried out here due to the very high computational cost. Also, note that the bandgaps, magnetic moments can also depend on the selection of pseudopotentials. Fully accurate *ab initio* calculations of 2D magnetic materials remain very challenging, even for a single material, with techniques like dynamical mean field theory (DMFT) and quantum Monte-Carlo (QMC) as possible approaches for future work.

Table. 2 *Comparison of PBE+SOC and $G_0W_0$+SOC bandgaps.*

| $E_g$ (eV) | PBE+SOC | $G_0W_0$+SOC |
|---|---|---|
| ZrFeCl$_6$ | 0.011 | 1.302 |
| VAg(PSe$_3$)$_2$ | 0.018 | 0.103 |
| RuCl$_3$ | 0.0 | 0.50 |

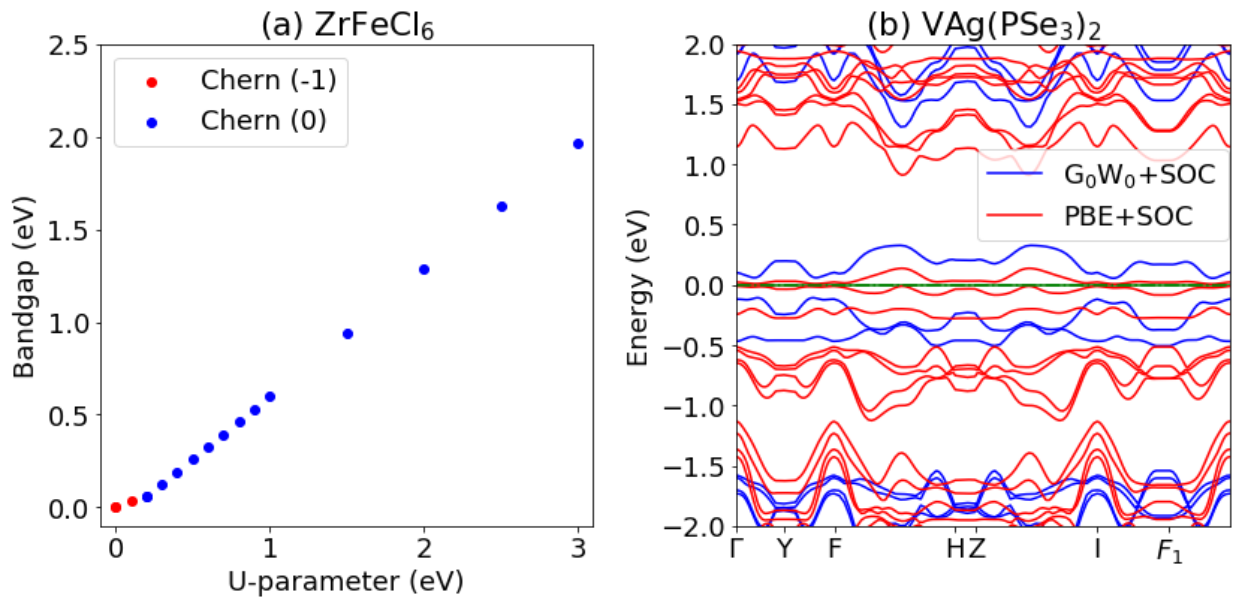

Fig. 6 *Effect of beyond DFT methods on the electronic structure using DFT+U and $G_0W_0$ methods. a) DFT+U effect on Chern number for ZrFeCl$_6$, b) bandstructure of VAg(PSe$_3$)$_2$ without (red) and with $G_0W_0$+SOC (blue) method.*

# 3 Conclusions

We have presented a comprehensive search of 2D topological materials, including both magnetic and non-magnetic materials, and considering insulating and metallic phases. Using the JARVIS-DFT 2D material dataset, we first identify materials with high spin-orbit spillage, resulting in 122 materials. Then we use Wannier-interpolation to carry-out $Z_2$, Chern-number, anomalous Hall conductivity, Curie temperature, and surface and edge state calculations to identify topological insulators and semimetals. For a subset of predicted QAHI materials, we run $G_0W_0$+SOC and GGA+U calculations. We find that as we introduce many-body effects, only a few materials retain a non-trivial band-topology, suggesting that higher-level electronic structure methods will be necessary to fully analyze the topological properties of 2D materials. However, we believe that as an initial step, the automated spillage screening and Wannier-approach provides useful predictions for finding new topological materials.

# 4 Methods

DFT calculations were carried out using the Vienna Ab-initio simulation package (VASP)[61,62] software using the workflow given on our github page (https://github.com/usnistgov/jarvis ). Please note commercial software is identified to specify procedures. Such identification does not imply recommendation by National Institute of Standards and Technology (NIST). We use the OptB88vdW functional[63], which gives accurate lattice parameters for both vdW and non-vdW (3D-bulk) solids[43]. We optimize the crystal-structures of the bulk and monolayer phases using VASP with OptB88vdW. The initial screening step for <1.5 eV bandgap materials is done with OptB88vdW bandgaps from the JARVIS-DFT database. Because SOC is not currently implemented for OptB88vdW in VASP, we carry out spin-polarized PBE and spin-orbit PBE calculations in order to calculate the spillage for each material. Such an approach has been validated by Refs. [8,64]. The crystal structure was optimized until the forces on the ions were less than 0.01 eV/Å and energy less than $10^{-6}$ eV. We also use $G_0W_0$+SOC and DFT+U on selected structures. We use Wannier90[65] and Wannier-tools[66] to perform the Wannier-based evaluation of topological invariants.

First, out of 963 2D monolayer materials, we identify materials with bandgap < 1.5 eV, and heavy elements (atomic weight ≥ 65). We calculate the exfoliation energy of a 2D material as the difference in energy per atom for bulk and monolayer counterparts[43]. Then we use the spillage technique to quickly narrow down the number of possible materials. As introduced in Ref.[42], we calculate the spin-orbit spillage, $\eta(\mathbf{k})$, given by the following equation:

$$\eta(\mathbf{k}) = n_{occ}(\mathbf{k}) - \text{Tr}(P\tilde{P}) \tag{1}$$

where,

$P(\mathbf{k}) = \sum_{n=1}^{n_{occ}(\mathbf{k})} |\psi_{n\mathbf{k}}\rangle\langle\psi_{n\mathbf{k}}|$ is the projector onto the occupied wavefunctions without SOC, and $\tilde{P}$ is the same projector with SOC for band $n$ and k-point $k$. We use a $k$-dependent occupancy $n_{occ}(\mathbf{k})$ of the non-spin-orbit calculation so that we can treat metals, which have varying number of occupied electrons at each k-point[8]. Here, 'Tr' denotes trace over the occupied bands. We can write the spillage equivalently as:

$$\eta(\mathbf{k}) = n_{occ}(\mathbf{k}) - \sum_{m,n=1}^{n_{occ}(\mathbf{k})} |M_{mn}(\mathbf{k})|^2 \tag{2}$$

where $M_{mn}(\mathbf{k}) = \langle \psi_{m\mathbf{k}} | \tilde{\psi}_{n\mathbf{k}} \rangle$ is the overlap between occupied Bloch functions with and without SOC at the same wave vector $\mathbf{k}$. If the SOC does not change the character of the occupied wavefunctions, the spillage will be near zero, while band inversion will result in a large spillage. In the case of insulating topological materials driven by spin-orbit based band inversion, the spillage will be at least 1.0 for Chern insulators or 2.0 for TRS-invariant topological insulators at the k-point(s) where band inversion occurs[42]. In other words, the spillage can be viewed as the number of band-inverted electrons at each k-point. For topological metals and semimetals, or cases where the inclusion of SOC opens a gap, the spillage is not required to be an integer, but we find empirically that a high spillage value is an indication of a change in the band structure due to SOC that can indicate topological behavior. We choose a threshold value of $\eta = 0.5$ at any k-point for our screening procedure, based on comparison with known topological semimetals[8]. Our screening method can also detect topological materials with small SOC, like the small band gap that is opened at the Dirac point in graphene[22] (JVASP-667). However, the screening method is not expected to work for semimetals with topological features that are not caused or modified by spin-orbit coupling.

After spillage calculations, we run Wannier based Chern and $Z_2$-index calculations for these materials.

The Chern number, $C$ is calculated over the Brillouin zone, BZ, as:

$$C = \frac{1}{2\pi}\sum_n \int d^2 \mathbf{k} \Omega_n \qquad (3)$$

$$\Omega_n(\mathbf{k}) = -\text{Im}\langle \nabla_k u_{nk} | \times | \nabla_k u_{nk} \rangle = \sum_{m \neq n} \frac{2\text{Im}\langle \psi_{nk}|\hat{v}_x|\psi_{mk}\rangle\langle \psi_{mk}|\hat{v}_y|\psi_{nk}\rangle}{(\omega_m - \omega_n)^2} \qquad (4)$$

Here, $\Omega_n$ is the Berry curvature, $u_{nk}$ being the periodic part of the Bloch wave in the *n*th band, $E_n = \hbar\omega_n$, $v_x$ and $v_y$ are velocity operators. The Berry curvature as a function of $\boldsymbol{k}$ is given by:

$$\Omega(\boldsymbol{k}) = \sum_n \int f_{nk}\Omega_n(\boldsymbol{k}) \tag{5}$$

Then, the intrinsic anomalous Hall conductivity (AHC) $\sigma_{xy}$ is given by:

$$\sigma_{xy} = -\frac{e^2}{\hbar}\int \frac{d^3k}{(2\pi)^3}\Omega(\boldsymbol{k}) \tag{6}$$

In addition to searching for gapped phases, we also search for Dirac and Weyl semimetals by numerically searching for band crossings between the highest occupied and lowest unoccupied band, using the algorithm from WannierTools[66]. This search for crossings can be performed efficiently because it takes advantage of Wannier-based band interpolation. In an ideal case, the band crossings will be the only points at the Fermi level; however, in most cases, we find additional trivial metallic states at the Fermi level. The surface spectrum was calculated by using the Wannier functions and the iterative Green's function method[67,68].

For magnetic systems, we primarily screen the ferromagnetic phase, with spins oriented out of the plane, which we expect is possible to achieve experimentally in most cases, using an external field if necessary. We initialize the magnetic moment with $6\mu_B$ for spin-polarized calculations. For a subset of interesting compounds, we perform a set of three additional calculations: ferromagnetic with spins in-plane, and antiferromagnetic with spins in- and out-of-plane. With these energies, we can fit a minimal magnetic model and calculate an estimated Curie temperature for ferromagnetic materials with out-of-plane spins, using the method of Ref.[54], which takes into account exchange constants and magnetic anisotropy. Magnetic anisotropy that favors out-of-plane spin is crucial to enable long-range magnetic ordering in two-dimensions following the Mermin-Wagner theorem[69]. We consider a simple Heisenberg model Hamiltonian with nearest-neighbor exchange interactions $J$, single-ion anisotropy $A$, and nearest neighbor anisotropic exchange $B$.:

$$H = -\frac{1}{2}\sum_{ij} J_{ij} \mathbf{S}_i \mathbf{S}_j - A \sum_j (S_i^z)^2 - \frac{1}{2}\sum_{ij} B_{ij} S_i^z S_j^z \qquad (7)$$

with $J_{ij}, A, B_{ij} > 0$. The sums run over all magnetic sites and $J_{ij} = J$, $B_{ij} = B$ if $i$ and $j$ are nearest neighbors and zero otherwise. The maximum value of $S_i^z$ denoted by S. For an Ising model, which corresponds to the limit of $A \to \infty$, the critical transition temperature is given by:

$$T_c^{Ising} = \frac{S^2 J \widetilde{T_c}}{k_B} \qquad (8)$$

where $\widetilde{T}_c$ is a dimensionless critical temperature with values of 1.52, 2.27, 2.27, and 3.64 for the honeycomb, quadratic, Kagomé, and hexagonal lattices respectively. However, systems with finite A, we instead use: [54]

$$T_c = T_c^{Ising} f\left(\frac{\Delta}{J(2S-1)}\right) \qquad (9)$$

with $\Delta = A(2S - 1) + BSN_{nn}$ and $f(x) = \tanh^{\frac{1}{4}}\left[\frac{6}{N_{nn}} log(1 + \gamma x)\right]$.

To better account for correlation effects and self-interaction error in describing transition metal elements, we apply the DFT+U method to a subset of materials[55,56]. For a few systems, we also carry out a systematic U-scan by varying the U parameter from 0 to 3 eV and monitor changes in the band structure[70]. We also perform $G_0W_0$+SOC[57-59] calculations with an ENCUTGW parameter (energy cutoff for response function) of 333.3 eV for a few materials to analyze the impact of many-body effects.

## 5 Data availability

The electronic structure data is made available at the JARVIS-DFT website: https://www.ctcms.nist.gov/~knc6/JVASP.html and http://jarvis.nist.gov .

# 6 Contributions

KC and KG developed the workflow. KC carried out the DFT calculations and analyzed the DFT data. KG carried out the magnetic ordering calculations. JJ and RP carried out the $G_0W_0$ based calculations. All contributed in writing the manuscript.

# 7 Competing interests

The authors declare no competing interests.

# 8 Supplementary information

Supplementary information for the spillage data is provided.